\documentclass[iop,twocolumn,amsmath,amssymb,superscriptaddress]{revtex4-1}
\usepackage{graphicx}
\usepackage{dcolumn}
\usepackage{bm}
\usepackage{natbib}
\usepackage{color}
\usepackage[normalem]{ulem}

\begin{document}

\title{Infrared phonons as a probe of a spin-liquid states in herbertsmithite ZnCu$_3$(OH)$_6$Cl$_2$} 

\author{A. B. Sushkov}
\affiliation{Center for Nanophysics and Advanced Materials, Department of Physics, University of Maryland, College Park, Maryland 20742, USA}
%\email{sushkov@umd.edu}
\author{G. S. Jenkins}
\affiliation{Center for Nanophysics and Advanced Materials, Department of Physics, University of Maryland, College Park, Maryland 20742, USA}
\author{Tian-Heng Han}
\affiliation{Materials Science Division, Argonne National Laboratory, Argonne, IL 60439}
\affiliation{James Franck Institute and Department of Physics, University of Chicago, Chicago, IL 60637}
\author{Young S. Lee}
\affiliation{Department of Applied Physics, Stanford University, Stanford, CA 94305}
\affiliation{Stanford Institute for Materials and Energy Sciences, SLAC National Accelerator Laboratory, 2575 Sand Hill Road, Menlo Park, CA 94025}
\author{H. D. Drew}
\affiliation{Center for Nanophysics and Advanced Materials, Department of Physics, University of Maryland, College Park, Maryland 20742, USA}

%\date{\today}

\begin{abstract}
We report on temperature dependence of the infrared reflectivity spectra of a single crystalline herbertsmithite in two polarizations --- parallel and perpendicular to the kagome plane of Cu atoms. 
We observe anomalous broadening of the low frequency phonons possibly caused by fluctuations in the exotic dynamical magnetic order of  the spin liquid.
 
\end{abstract}

\maketitle

ZnCu$_3$(OH)$_6$Cl$_2$, also known as herbertsmithite, is an example of a geometrically frustrated quantum spin liquid~\cite{Anderson_1973}.   
Herbertsmithite crystalizes in a lattice with kagome planes of highly frustrated corner sharing triangles of spin-1/2 Cu ions.  
It is a disordered antiferromagnet down to low temperatures $T = 50$~mK $\ll J$, where $J\approx200$~K is the exchange energy. 
Because quantum effects are enhanced for spin-1/2 materials there has been much theoretical discussion and experimental studies addressing the possible ground states~\cite{Sachdev_1992,Mambrini_2000,Nikolic_2003,Braithwaite2004,Shores2005,Bert_2007,Helton_2007,Lee_2007,deVries_2008,Hermele_2008,Imai_2008,Lee_2008,Olariu_2008,deVries_2009,Ofer_2009,Freedman_2010,Helton_2010,Liu_2010,Mendels_2010,Wen_2010,Wulferding_2010,Jeong_2011,Lu_2011,Mendels_2011,Yan_2011,Fak_2012,Han_2012,Han_2012n,deVries_2012,Shaginyan_2012,Capponi2013,Dodds_2013,Hao_2013,Jeschke_2013,Pilon_2013,Potter_2013,Pujari_2013,Rousochatzakis_2013,Serbyn_2013,Asaba_2014,Colbert_2014,Rousochatzakis_2014,Taillefumier_2014,Furukawa_2015,Guterding_2015,Han_2015,Hu_2015,Iqbal_2015,Iqbal_2015_2,Shaginyan_2015,Zhu_2015}. 
The basic questions are whether the spin excitations are gapped in the ground state and whether the singlet valence bonds are static (valence bond solid --- VBS)~\cite{Nikolic_2003} or dynamic (resonating valance bond --- RVB)~\cite{Mambrini_2000}.  
The RVB models come in gapped and gapless versions.  
Detecting a gap at low temperatures has been hindered by inadvertent magnetic disorder.  
However, recent experiments report on a spin gap $\Delta_s = 10$~K based on NMR measurements~\cite{Fu_2015}.

One of the theoretical RVB models  using contractor renormalization (CORE) method predicts a gapped state that breaks reflection symmetries and possess a chiral $p6$ symmetry~\cite{Capponi2013}.  
It has been noted that this symmetry can induce a splitting of the doubly degenerate in-kagome plane optical phonons through spin-phonon coupling~\cite{SP-1968,us-PRL-2005}. 
One of the theoretical VBS models assumes the ground state of herbertsmithite to be a gapless algebraic spin liquid and predicts possible contribution of the VBS order into lineshapes of optical phonons via spin-phonon coupling~\cite{Hermele_2008}. 
Therefore, there is interest in studies of the optical phonons for evidence of these predictions. 

Another motivation for studying infrared optical phonons comes from the recent results on another spin-liquid compound Tb$_2$Ti$_2$O$_7$~\cite{Fennell2014,Ruminy2016,Ruminy2016-2}. 
Inelastic neutron scattering experiments revealed the hybridization of Tb$^{3+}$ crystal-field excitons and transverse acoustic phonons which suppresses both magnetic ordering and the structural distortion~\cite{Fennell2014}. 
Disorder can induce magnetic ordering in this compound but the hybridization is not suppressed by disorder~\cite{Ruminy2016}. 
Comparative study of Tb, Dy, and Ho titanate pyrochlores by first-principles calculations and inelastic neutron and x-ray scattering experiments produced a lot of information about phonon spectra which is important in understanding of spin-lattice interaction and formation of the spin-liquid and spin-ice states~\cite{Ruminy2016-2}. 
We are looking for signatures of magnetic fluctuations in infrared-active phonons of herbertsmithite. 

Herbertsmithite~\cite{Braithwaite2004} was first found in 1972 in Chile, synthesized in 2005~\cite{Shores2005}, and high-quality single crystals of millimeter size have been available since 2011~\cite{Han2011}. 
Neutron and synchrotron x-ray diffraction experiments on synthetic single crystals~\cite{Han2011} found no obvious structural transition down to 2~K.  
In this paper, we report the temperature dependence of the infrared optical phonon spectra of a herbertsmithite single crystal.  
Our data show anomalous broadening of the low frequency infrared phonons upon cooling.  

Single crystals of ZnCu$_3$(OH)$_6$Cl$_2$ were grown as described elsewhere~\cite{Han2011}.
Fourier transform infrared reflectivity measurements were performed using a 2 mm in diameter aperture on the largest face of the single crystal. 
From polarization dependence of the reflectivity spectra, we found two orthogonal polarizations with distinct phonon resonances. 
We do not see mixture of the contributions between polarizations. 
This means that $c$-axis is sufficiently close to the face of the crystal for phonon study accuracy.

We determine the number of allowed infrared active phonons using the Correlation method~\cite{Fateley1972}. 
Space group of herbertsmithite crystal is $R\bar{3}m$ with three formula units in hexagonal unit cell\cite{Shores2005,Colman2011,Malcherek2014} which is reflected in the 2nd column of Table~1.
However, the primitive unit cell is rhombohedral and it contains one formula unit which is reflected in the 3rd column of Table~1 where we divided the number of Wyckoff sites by 3. 
Thus, the total number of phonons is $3N=54$, where $N=18$ is the number of atoms in the primitive unit cell.  
Presence of two types of the unit cell settings in literature does not affect phonon calculation because Wyckoff sites $a$ and $b$ have the same site symmetry which is also valid for sites $d$ and $e$.  
\begin{table*}
\caption{\label{tab1}Allowed $\Gamma$-point phonons for herbertsmithite by Correlation method~\cite{Fateley1972}. 
Space group $R\bar{3}m$ (\#166), one formula unit ZnCu$_3$(OH)$_6$Cl$_2$ per rhombohedral primitive unit cell.}
\begin{ruledtabular}
\begin{tabular}{ccccc}
 Species & Wyckoff   & For phonon  & Site     & Vibrational \\
 & positions & calculation & symmetry & modes     \\
\hline
Zn & 3a\footnotemark[1] or 3b\footnotemark[2] & a or b & $D_{3d}$ & $A_{2u} + E_{u}$ \\
Cl & 6c & 2c & $C_{3v}$ & $A_{1g} + A_{2u} + E_g + E_{u}$ \\
Cu & 9d\footnotemark[1] or 9e\footnotemark[2] & 3d or 3e & $C_{2h}$ & $A_{1u} + 2A_{2u} + 3E_{u}$ \\
O & 18h & 6h & $C_s$ & $2A_{1g} + A_{1u} + A_{2g} + 2A_{2u} + 3E_g + 3E_{u}$ \\
H & 18h & 6h & $C_s$ & $2A_{1g} + A_{1u} + A_{2g} + 2A_{2u} + 3E_g + 3E_{u}$ \\
\hline
\multicolumn{3}{c}{ } & \multicolumn{1}{r}{Total:} & \multicolumn{1}{c}{$5A_{1g} + 3A_{1u} + 2A_{2g} + 8A_{2u} + 7E_g + 11E_{u}$} \\
\hline
\multicolumn{3}{c}{ } & \multicolumn{1}{r}{Acoustic:} & \multicolumn{1}{c}{$A_{2u} + E_{u}$} \\
\multicolumn{3}{c}{ } & \multicolumn{1}{r}{Infrared:} & \multicolumn{1}{c}{$7A_{2u}(e\parallel c) + 10E_{u}(e\perp c)$} \\
\multicolumn{3}{c}{ } & \multicolumn{1}{r}{Raman:} & \multicolumn{1}{c}{$5A_{1g}(\alpha_{xx} + \alpha_{yy}, \alpha_z) + 7E_g(\alpha_{xx} - \alpha_{yy}, \alpha_{xy}), (\alpha_{xz},\alpha_{yz})$} \\
\multicolumn{3}{c}{ } & \multicolumn{1}{r}{Silent:} & \multicolumn{1}{c}{$3A_{1u} + 2A_{2g}$} \\
\end{tabular}
\end{ruledtabular}
\footnotetext[1]{From Ref.~\onlinecite{Shores2005,Colman2011}.}
\footnotetext[2]{From Ref.~\onlinecite{Malcherek2014}.}
\end{table*} 
Thus, for infrared active phonons, we expect to observe seven nondegenerate $A_{2u}$ modes in $e||c$ polarization, where $e$ is electric field of light and $c$-axis is perpendicular to the kagome plane, and 10 doubly degenerate $E_u$ modes polarized in the kagome plane. 
Raman-active phonon analysis is identical to the published earlier results~\cite{Wulferding_2010,deVries_2012}.    

The number of phonons observed in our reflectivity spectra is larger than predicted for the $R\bar{3}m$ group. 
We do not consider the possibility of another space group because all structural studies result in one space group. 
The most probable origin of additional vibrational modes is crystal defects such as Cu/Zn counterdoping, vacancies, impurities. 
In general, any substitution ion in a lattice which differs by mass or spring constant results in new, localized modes~\cite{Barker_Sievers_1975}. 
Structural studies of our crystals show that the kagome plane consists of Cu ions only and 5 to 10~\% of Cu reside in Zn plane\cite{Han_2012,Han_2012n}. 
We do not study additional phonon modes here, instead, we focus on magnetic effects on phonons. 
The presence of additional modes does not affect our main result --- the observation of unusual broadening of some phonons upon cooling. 
Both eigenmodes of an ideal crystal and additional local modes get narrower upon cooling. 
Only magnetic interactions in this compound can cause such broadening of phonons. 

Figure~1 shows the measured reflectivity spectra of herbertsmithite together with the fit results over a broad frequency range. 
Modes seen in the frequency range approximately 100--3600~cm$^{-1}$ are phonons. 
Qualitatively, we can say that low frequency phonons are dominated by heaviest ions --- Cu and Zn in this case, mid range phonons, say $a6$-$a9$ and $c3$-$c6$, are oxygen and chlorine dominated, and the highest modes $a10$ and $c7$ are hydrogen dominated. 
Magnetic effects are expected at low frequency phonons dominated by magnetic Cu$^{2+}$ ions.  
As the number of observed phonons is larger than predicted for the given space group, the assignment of phonons is somewhat arbitrary. 
We assign the strongest modes using the room temperature spectra where magnetic effects are minimal.  
We measured temperature dependence of reflectivity in two polarizations below 6,000~cm$^{-1}$. 
Reflectivity spectra above 6,000~cm$^{-1}$ were measured at room temperature without polarizer. 
A peak in the optical conductivity at 32,000~cm$^{-1}$ (4~eV) is the optical gap. 
\begin{figure}
\includegraphics[width=\columnwidth]{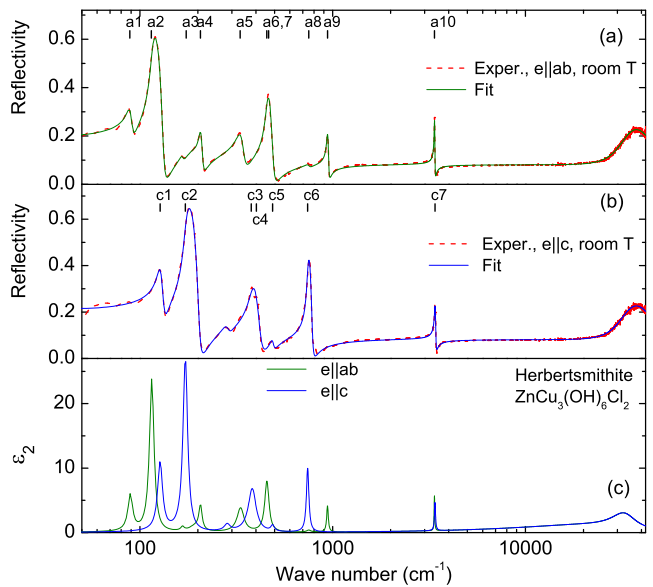}
\caption{(Color online). 
(a),(b) Room temperature reflectivity spectra measured and obtained from fits of the Lorentzian model (Eq.~(\ref{eq:lorentz})). 
Electric field of light $e$ is (a) in the kagome plane of Cu atoms, (b) perpendicular to the kagome plane.  
(c) Imaginary part of the dielectric constant obtained from fits for two polarizations. 
A peak at 32,000~cm$^{-1}$ (4~eV) is the optical gap.  
Resonances below 3,600~cm$^{-1}$ are phonons. 
$a1$-$a10$ and $c1$-$c7$ are tentative assignments of phonons.
} 
\label{R}
\end{figure}

Figure~2 shows temperature dependence of the measured reflectivity spectra of herbertsmithite in the bolometer frequency range. 
Phonon modes are usually sensitive to temperature and this property can help in identification of phonons among observed features in the reflectivity spectra. 
We assign the modes in panel (a) near 500~cm$^{-1}$ as two major phonons $a6$ and $a7$ which become clear at low temperatures. 
In spite of its complicate line shape, we consider mode $c2$ in panel (b) as one phonon and we fit it with one oscillator because it is expected to be a phonon singlet according to the symmetry analysis. 
Even without looking at the fit results, one can tell that phonons $a2$ and $c1$ show temperature behavior different from other phonons. 
Upon cooling, the reflectivity peaks of all other phonons grow while $a2$ and $c1$ phonons have the opposite trend because of broadening which is possibly of magnetic origin. 

We do not assign broad features between 250 and 350~cm$^{-1}$ in Fig.~2(b) as phonons because they don't depend on temperature and they are too broad to be phonons. 
This can be a low energy electronic transition but it should depend on temperature in this frequency range as well. 
We note that it is seen only in $e||c$ polarization. 
An asymmetric mode with Fano line shape and $A_{1g}$ symmetry was observed at 230~cm$^{-1}$ in Raman spectra~\cite{Wulferding_2010,deVries_2012}. 
It was assigned as an additional lattice mode due to crystallographic disorder. 
Its Fano shape was explained as the result of interaction with a continuum of states, possibly, of spin fluctuations. 
Possibly, we are seeing that continuum in our infrared spectra. 
Another possibility --- we observe this continuum as the electric dipole active excitation which means the first order coupling between spin fluctuations and phonons similar to the electromagnon effect in noncollinearly ordered magnets\cite{Sushkov_Mostovoy_2008}.

%\begin{figure*}
\begin{figure}
\includegraphics[width=\columnwidth]{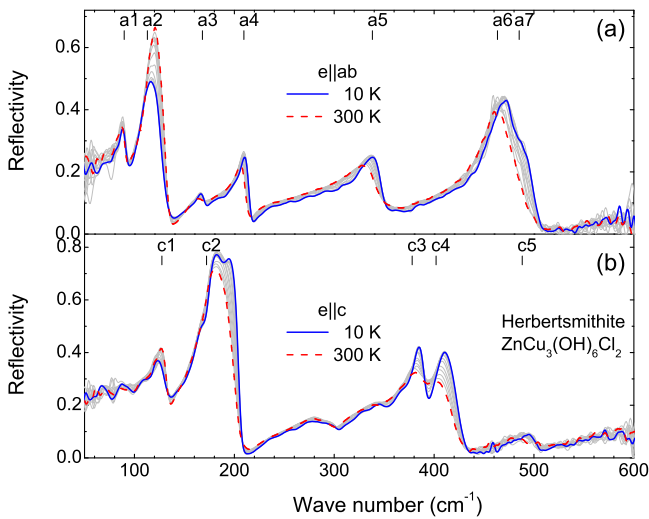}
\caption{(Color online). Temperature dependence of the reflectivity spectra of a herbertsmithite single crystal in the 4~K bolometer frequency range where magnetic effects are most expected. 
Grey lines are spectra at intermediate temperatures. 
} 
\label{bolo}
\end{figure}

Figure~3 is similar to Fig.~2 and it presents higher frequency phonons. 
We see extra phonons here in both polarizations. 
Group of phonons $c7$ shows nonmonotonic temperature dependence. 
In two spectra taken at 20 and 30~K, they are stronger than in the 10~K spectrum. 
Not likely that this is an experimental error. 
The two bottom panels show different frequency ranges of the same spectra. 
Thus, the nonmonotonic behavior of $c7$ phonon group is real but not understood.  

 We extract phonon parameters by fitting the reflectivity spectra with the model of a sum of Lorentzian oscillators of the Reffit program~\cite{Kuzmenko_2005}. 
The complex dielectric function $\varepsilon=\varepsilon_1+i\varepsilon_2$ takes the form:
\begin{equation}
    \varepsilon(\omega)=\varepsilon_{\infty}+
    \sum_j\frac{\omega_{sj}^2}{\omega_{0j}^2-\omega^2-i\omega\gamma_j} .
\label{eq:lorentz}
\end{equation}
Here $\omega_{0j}$, $\gamma_j$, and $\omega_{sj}$ are the resonance, the damping, and the spectral weight  frequencies of the $j$th electric dipole active mode, respectively. 
The complex optical conductivity is related to the dielectric function by $\sigma=\sigma_1+i\sigma_2 = \omega\varepsilon/4\pi i$.

\begin{figure}
\includegraphics[width=\columnwidth]{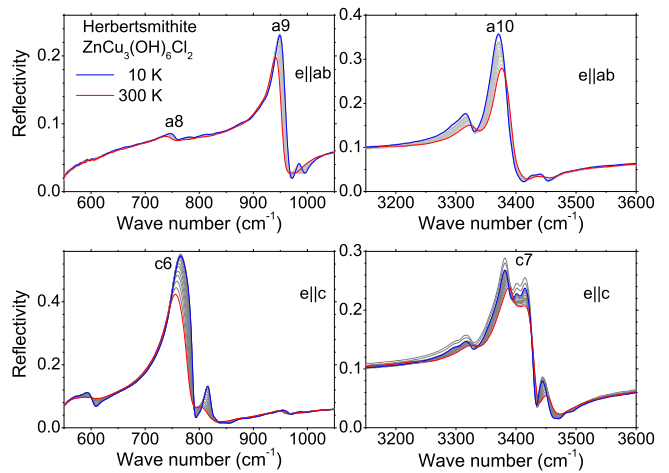}
\caption{(Color online). Temperature dependence of the reflectivity spectra in the MCT detector frequency range. 
Grey lines are spectra at intermediate temperatures.} 
\label{MCT}
\end{figure}
\begin{figure*}
\includegraphics[width=7.3 in]{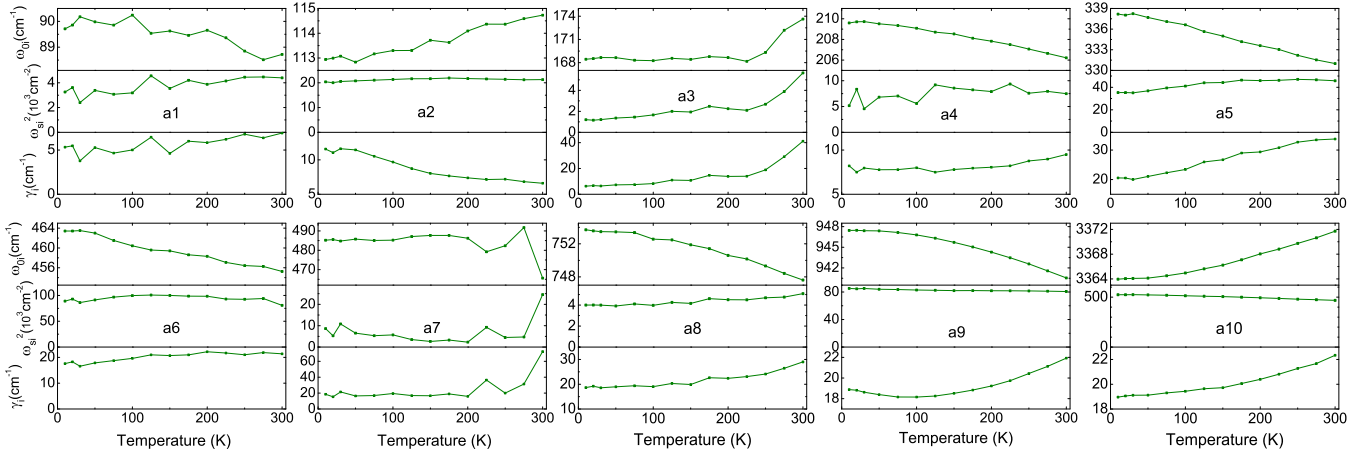}
\caption{(Color online). Temperature dependence of the phonon fit parameters determined by Eq.~(\ref{eq:lorentz}) for ten major phonons polarized in the kagome plane. 
Parameters of extra phonons are shown in Supplemental material~\cite{Suppl}.
} 
\label{Pa}
\end{figure*}

Figure~4 shows phonon fit parameters versus temperature for in-plane polarization. 
Usually, phonon parameters exhibit the following trend upon cooling. 
The resonance frequencies $\omega_{0j}$ harden, the spectral weights $\omega_{sj}^2$ stay constant or grow slightly, and the scattering rates $\gamma_j$ decrease. 
$a5$ phonon may be a good example of such temperature behavior. 
Most of the other nine phonons behave in a similar way. 
Softening of the $a2$ mode may also be a spin-phonon coupling effect but the $a10$ hydrogen dominated mode also shows softening which is unlikely to be magnetic in origin. 
The $a2$ phonon also shows very unusual broadening upon cooling which we can only understand as magnetic in origin. 
Splitting of a phonon doublet into two singlets by specific chiral magnetic ordering of Cu spins was predicted theoretically for a kagome plane of spins~\cite{Capponi2013}. 
Phonon line shape looks like a single mode and fitting of the $a2$ phonon line with two oscillators close in frequency results in rejection of one of the two oscillators by the Reffit program. 
Thus, we cannot detect possible splitting of the $a2$ phonon although 4 cm$^{-1}$ spectral resolution of the measured spectra was sufficient for that task. 
This absence of detectable splitting does not rule out the whole model because the predicted splitting becomes zero at certain range of the exchange parameters. 
For a VBS model, it was shown that competing VBS orders may contribute to all three optical phonon parameters~\cite{Hermele_2008}. 
The sign of the predicted frequency shift is unknown, the spectral weight may also change due to effective charge redistribution, and only additional broadening is the unambiguous prediction which we observe in the $a2$ mode temperature dependence. 
Phonon $a1$ shows quite small narrowing toward zero temperature which is surprising for such a low frequency phonon and also may be a sign of magnetic broadening competing with thermal narrowing of the phonon line.  
Phonon $a9$ surprisingly broadens below 100~K and we cannot explain such a behavior. 
First-principle calculations of phonons for herbertsmithite may give us better understanding of the observed behavior by separating lattice effects from magneto-lattice ones. 
 
Figure~5 shows the phonon fit parameters versus temperature for the polarization perpendicular to the kagome plane in bolometer frequency range. 
Here, phonons $c1$ and $c4$ get slightly broader and phonon $c2$ gets slightly narrower upon cooling which may be understood as presence of magnetic broadening especially in comparison with phonon $c3$. 
Fit results on phonon width of $c5$ are not reliable at low temperatures as it is seen from the spectra in this figure. 
\begin{figure*}
\includegraphics[width=7.3 in]{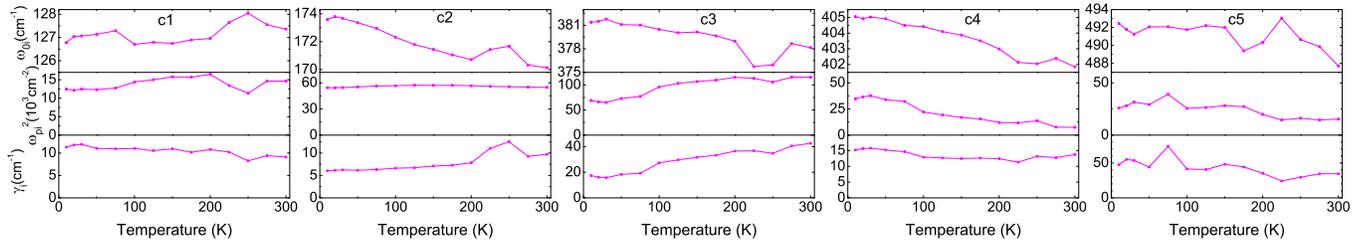}
\caption{(Color online). Temperature dependence of the phonon fit parameters determined by Eq.~(1) for five low frequency phonons polarized perpendicular to the kagome plane. 
Parameters of higher frequency phonons are shown in Supplemental material~\cite{Suppl}. } 
\label{Pc}
\end{figure*}

Spin liquid on a kagome lattice posses different types of spin dynamics in various temperature ranges~\cite{Taillefumier_2014} characterized by the ratio $k_BT/J$. 
In our temperature range, we possibly deal with paramagnetic regime at $T>J=200$~K and with dipolar-like spin correlations below 200~K. 
Spin excitations can be probed by infrared spectroscopy directly at the frequencies of the spin excitations as magnetic dipole active or electric dipole (ED) active modes. 
To acquire electric dipole activity essentially magnetic spin excitations have to borrow ED activity from ED-allowed transitions, most probably from the infrared optical phonons --- the effect known as electromagnon which was observed in some ordered frustrated magnets\cite{Pimenov-Nature,us-PRL-2007,Kida-2009}. 
Spin orders can couple to phonons in the second order. 
This effect is called spin-phonon coupling~\cite{SP-1968,us-PRL-2005} and it leads to a magnetic shift (positive or negative) of the phonon frequency. 
In magnetically frustrated cubic spinel ZnCr$_2$O$_4$, accumulated magnetic energy leads eventually to a magneto-structural phase transition at 12~K. 
In the low temperature phase, the crystal structure becomes tetragonal, one triply degenerate phonon splits into phonon doublet and singlet, and frequency splitting is proportional to the magnetic energy~\cite{us-PRL-2005}. 
In herbertsmithite, there is no structural transition and we don't expect clear phonon splitting but rather broadening of a phonon line. 
Interestingly, the broadening of the $a2$ phonon is about 10~\% of its resonance frequency as seen from Fig.~4 and phonon splitting predicted in~\cite{Capponi2013} is also 10~\% of the resonance frequency. 

We believe that the broadening effect on the $a2$ phonon is magnetic in origin and it has to do with valence-bond physics. 
At low temperature, the system may be settling into a spin-liquid regime, in which valence-bond singlets formed by neighboring Cu spins slowly move around. 
Bonds with and without singlets will tend to have different lengths and spring constants --- the same effect as in ZnCr$_2$O$_4$~\cite{us-PRL-2005}. 
Thus a phonon will find that the lattice is a bit disordered and the phonon line will exhibit some broadening. 
{\it Ab initio} calculations similar to Ref.~\onlinecite{Fennie2006} would be very useful for further understanding of the spin-lattice dynamics in this material. 

In conclusion, we report possible observation of magnetic broadening of at least one phonon mode in a herbertsmithite single crystal. 
This observation is in agreement with the prediction of one of the VBS models~\cite{Hermele_2008}. 
Further experimental and theoretical studies are needed to understand the origin of the magnetic effects on phonons, and to separate the eigenmodes of the $R\bar{3}m$ space group from the observed additional vibrational modes. 

\acknowledgments
This work was supported by DOE under grant \# ER 46741-SC0005436. 
Some part of the work was performed at the Aspen Center for Physics, partially funded by NSF Grant No. 1066293.  
We thank Assa Auerbach, Oleg Tchernyshyov, and Karen Rabe for inspiring discussions.

\end{document}